\documentclass[seceq]{ptptex}

\usepackage{graphicx}
\usepackage{wrapft}

\def\be{\begin{equation}}
\def\ee{\end{equation}}
\def\bea{\begin{eqnarray}}
\def\eea{\end{eqnarray}}
\def\la{\langle}
\def\ra{\rangle}
\def\dis{\displaystyle}
\def\muh{\hat{\mu}}
\def\bra{\langle}
\def\ket{\rangle}

\title{
Hadronic property at finite density
}

\author{
Tetsuya  
\textsc{Takaishi}%
}

\inst{
Hiroshima University of Economics, Hiroshima 731-0192, Japan
}

\abst{
We report on three topics on finite density simulations:
(i) the derivative method for hadronic quantities,
(ii) phase fluctuations in the vicinity of the critical temperature
and
(iii) the density of states method at finite isospin density.
}

\begin{document}

\maketitle

\section{Introduction}
Lattice QCD simulations at finite chemical potential $\mu$ 
are extremely difficult due to the sign problem.
Recently it has been realized that 
at small chemical potential 
one can study density effects on physical quantities 
by various approaches\cite{Review1}.
One of the approaches is the derivative method which 
has been used for the study of response of meson masses 
with respect to $\mu$ by QCD-TARO collaboration\cite{QCD-TARO,QCD-TARO2}.
The original idea of the derivative method 
may date back to the study of the fermion number susceptibilities\cite{SUS}
where the derivative of the fermion number density with repect to 
$\mu$ were calculated.
In section 2, we report on results from the derivative method 
for the response of meson masses as well as the chiral condensate.

At large $\mu$ we believe that the standard Monte Calro method based on importance sampling fails
due to the sign problem, 
i.e. the phase fluctuation is expected to be large for large $\mu$. 
However, we do not know exactly how the phase fluctuates with various simulation parameters.
In section 3, we give results of the phase fluctuation in the vicinity of 
the critical temperature. We also present results of 
$\la\bar{\psi}\psi\ra$ and the Polyakov loop for $\mu \le 0.25$. 

The most lattice simulations have been performed by methods based on importance sampling.
There exists an alternative method called density of states method.
An advantage of the density of states method is that 
one can obtain results for various values of simulation parameters 
without making independent simulations
although the lattice size may be limited to a small one.  
In section 4, we give results from the density of states method for isospin density.
Since the system at isospin density has no sign problem 
the results are compared with those from the standard Monte Carlo method.

\section{Derivative method for hadronic quantities}

The derivative method extracts physical information
at a small chemical potential without 
directly simulating a system at finite chemical potential.
The original idea of the derivative method comes 
from the calculations of the fermion number susceptibilities\cite{SUS}.
The QCD-TARO collaboration\cite{QCD-TARO,QCD-TARO2} studied the response of meson masses
by the derivative method. 
The method was also used for the study of the pressure\cite{Gavai2003,Allton2003}.

Let us consider an observable $O(\mu)$. 
The expectation value of the observable $O(\mu)$ is given by
\be 
<O>=\frac{\int O \Delta e^{-S_g}dU}{\int \Delta e^{-S_g}dU},
\ee
where $S_g$ is the gluonic action and $\Delta$ stands for the fermion determinant.
For two flavors of staggered quarks ({\it u} and {\it d} quarks), $\Delta$ is written as
\bea
\Delta & = & \det M(\mu_u)^{1/4} \det M(\mu_d)^{1/4} \\
& = & \exp(\frac14 Trln M(\mu_u)+ \frac14 Trln M(\mu_d)).
\eea
The first derivative of $\la O\ra$ with respect to $\mu$ at $\mu=0$ is given by
\be
\frac{\partial \la O\ra}{\partial \muh} =\la\dot{O}+O\frac{\dot{\Delta}}{\Delta}\ra,
\label{deriv1}
\ee
where we used $\dis\dot{O}=\frac{\partial O}{\partial \mu}$ for simplicity.
Note that here we used 
$\dis\la\frac{\dot{\Delta}}{\Delta}\ra=0$ at $\mu=0$.
Similarly the second derivative at $\mu=0$ is given by
\be
\frac{\partial^2 <O>}{\partial \mu^2} =<\ddot{O}+2\dot{O}\frac{\dot{\Delta}}{\Delta}>
-<O\circ\frac{\ddot{\Delta}}{\Delta}>_{cc},
\label{deriv2}
\ee
where $<A\circ B>_{cc}= <AB>-<A><B>$.

Next let us consider the spatial hadronic correlator,
\be  C(x)\equiv \sum_{y,z,t}\la H(x,y,z,t)H(0,0,0,0)^\dagger\ra,
\ee
and take derivatives of $C(x)$.
We assume that $C(x)$ is dominated by a single pole contribution,
\be 
C(x)=A(e^{-Mx}+e^{-M(L-x)}),
\ee
where $M$ is the hadron mass and 
$L$ is the lattice size in the $x$ direction.
Taking the first and the second derivatives of $C(x)$ 
with respect to $\mu$ we obtain
\be
\frac1{C(x)}\frac{dC(x)}{d\mu}=\frac1{A}\frac{dA}{d\mu}+\frac{dM}{d\mu}\times
\{(x-\frac{L}{2})\tanh[M(x-\frac{L}{2})]-\frac{L}{2}\},
\ee
and 
\begin{flushleft}
\bea
\frac1{C(x)}\frac{d^2C(x)}{d\mu^2}& = & \frac1{A}\frac{d^2A}{d\mu^2}+(\frac2{A}\frac{dA}{d\mu}\frac{dM}{d\mu}+\frac{d^2M}{d\mu^2})
 \{(x-\frac{L}{2})\tanh[M(x-\frac{L}{2})]-\frac{L}{2}\} \\ \nonumber
                                  & + & (\frac{dM}{d\mu})^2\{(x-\frac{L}{2})^2+\frac{L^2}{4}
-L(x-\frac{L}{2})\tanh[M(x-\frac{L}{2})]\}.
\eea
\end{flushleft}
The left hand sides of these equations are used as fitting functions to the Monte Carlo data
given by eq.(\ref{deriv1}) and (\ref{deriv2}). 
The response of masses with respect to $\mu$, i.e. $\dis \frac{dM}{d\mu}$ and $\dis \frac{d^2M}{d\mu^2}$
are given as the fitting parameters.

\begin{wrapfigure}{r}{7.6cm}
    \vspace{5mm}
       \centerline{\includegraphics[width=7.6 cm]
                                   {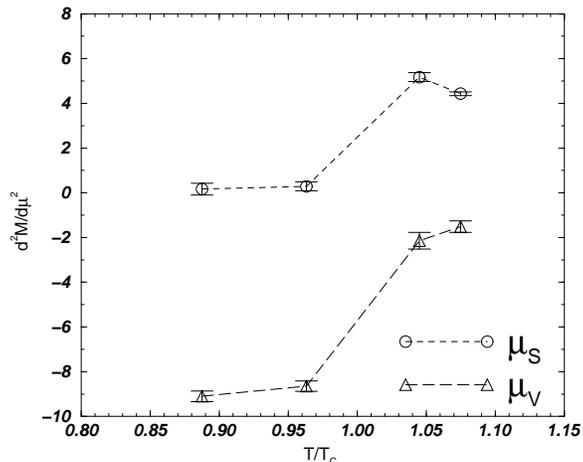}}
   \caption{Second responses $d^2M/d\mu_S^2$ and  $d^2M/d\mu_V^2$
of the PS meson mass at $m_q a= 0.025$ as a function of $T/T_c$\cite{QCD-TARO,QCD-TARO2}. }
   \label{fig:SECONDMSS}
\vspace{6mm}
\end{wrapfigure}

The QCD-TARO collaboration studied the response of 
the pseudo scalar (PS) meson screening mass for $n_f=2$ flavors 
on $16\times 8^2\times 4$ lattices.
The first response of the PS meson mass turned out to be consistent with zero.
Thus the first relevant response is the second one.
Figure~\ref{fig:SECONDMSS} shows the second responses of the PS meson mass.
Here the isoscalar and isovector chemical potentials  are defined by
\be
\mu_S=\mu_u=\mu_d,
\ee
and
\be
\mu_V=\mu_u=-\mu_d,
\ee
respectively. The isovector chemical potential is also called 
isospin chemical potential 
and we also use $\mu_V$ to stand for the isospin chemical potential. 

In the low temperature phase, the dependence of the mass on $\mu_S$ is
small. This behavior is to be expected since below $T_c$
the PS meson is a Goldstone boson and persists its zero mass feature. 
On the other hand, above $T_c$ the PS meson loses 
the Goldstone nature and can obtain its mass,
as a result $d^2M/d\mu_S^2$ seems to remain finite.

The PS meson on $\mu_V$ seems to behave differently.
In the low temperature phase the PS meson mass tends to decrease 
with $\mu_V$. In the high temperature phase 
the PS meson also seems to decrease but 
the rate of the decrease is small.

Next we consider the derivatives of $\la\bar{\psi}\psi\ra$.
At $\mu=0$  the first derivatives of $\la\bar{\psi}\psi\ra$ with respect to
both isoscalar and isovector chemical potentials are identically zero, i.e.
\be
\frac{\partial \la\bar{\psi}\psi\ra}{\partial \mu_S}=\frac{\partial \la\bar{\psi}\psi\ra}{\partial \mu_V}=0.
\ee
   \begin{figure}
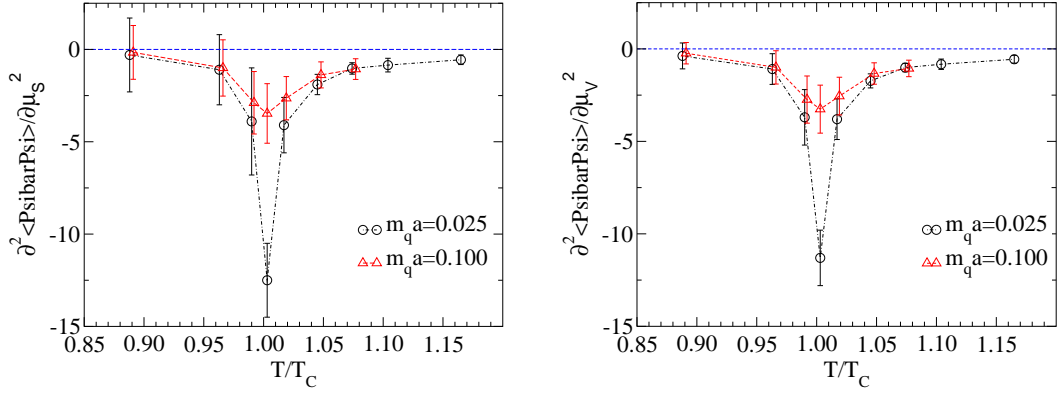

\vspace{1cm}
       \centerline{\includegraphics[width=6.6 cm]
                                   {dpsidmuS.eps}
\hspace{5mm}
\includegraphics[width=6.6 cm]
                                   {dpsidmuV.eps}}
\caption{$\dis\frac{\partial^2 \la \bar{\psi}\psi\ra}{\partial \mu_S^2}$ (left) and
$\dis\frac{\partial^2 \la \bar{\psi}\psi\ra}{\partial \mu_V^2}$ (right)
as a function of $T/T_c$\cite{PSIBARPSI}.}
   \label{fig:PSIBARPSI}
   \end{figure}
Thus the first relevant term is the second derivative.
Figure~\ref{fig:PSIBARPSI} shows the second derivatives 
of $\la\bar{\psi}\psi\ra$ with respect to $\mu_S$ and $\mu_V$ for $n_f=2$ flavors on
$16\times 8^2 \times 4$ lattices, 
calculated by Choe {\it et al.}\cite{QCD-TARO,PSIBARPSI}.
No significant difference can be seen between $\dis\frac{\partial^2 \la \bar{\psi}\psi\ra}{\partial \mu_S^2}$ 
and $\dis\frac{\partial^2 \la \bar{\psi}\psi\ra}{\partial \mu_V^2}$.
The strength of the responses increases as the quark mass decreases.
The similar results are also obtained for the NJL model\cite{NJL}.

The results of the second derivatives can be used for 
evaluating $\la\bar{\psi}\psi\ra$ at small chemical potentials as
\be
\la \bar{\psi}\psi(\mu)\ra \approx \la \bar{\psi}\psi(0)\ra +
\frac12\frac{\partial^2 \la \bar{\psi}\psi\ra}{\partial \mu^2} \mu^2.
\label{pol1}
\ee
\begin{figure}
    \vspace{5mm}
       \centerline{\includegraphics[width=7.6 cm]
                                   {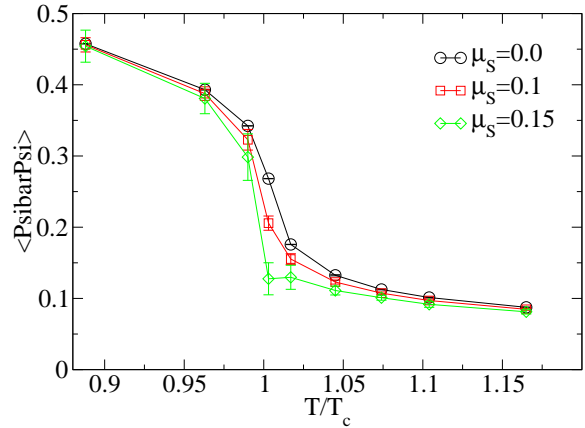}}
   \caption{$\la\bar{\psi}\psi\ra$ extrapolated to finite $\mu_S$ at $m_q=0.025$\cite{PSIBARPSI}.
Here $T_c$ is the critical temperature at $\mu=0$.}
   \label{fig:psifinite}
\end{figure}
Figure~\ref{fig:psifinite} shows $\la\bar{\psi}\psi(\mu_S)\ra$ 
extrapolated to finite $\mu_S$ using eq.(\ref{pol1}).
We see that qualitatively the critical temperature decreases as
$\mu$ increases. Similar behavior can be seen for the case of 
the isovector chemical potential.

\section{Phase fluctuation in the vicinity of the critical temperature}
\begin{wrapfigure}{r}{7.6cm}
       \centerline{\includegraphics[width=7.6 cm]
                                   {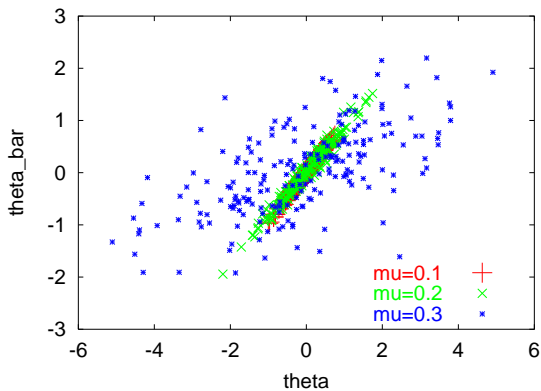}}
   \caption{$\bar{\theta}$ vs the exact phase $\theta$ for 3 values of $\mu$\cite{FKT}.}
   \label{fig:thetabar}
\end{wrapfigure}

There is no satisfactory method to simulate the system at large $\mu$.
In principle the reweighting method can be applied for any $\mu$ as
\be
\la O(\mu) \ra = \frac{\la O(\mu)e^{i\theta}\ra_{\mu_V}}{\la e^{i\theta}\ra_{\mu_V}},
\label{reweight}
\ee 
where $\dis \la O \ra_{\mu_V}$ stands for the expectation value of the operator $O$ 
in an ensemble at isospin density, i.e. the configurations are generated with the phase quenching measure 
$\sim|\det M(\mu)|^{n_f/4}e^{-S_g}$. 
For large $\mu$ eq.(\ref{reweight}) is expected to be impractical
since $\la e^{i\theta}\ra_{\mu_V}$ becomes small.
For such small $\la e^{i\theta}\ra_{\mu_V}$, in order to obtain a meaningful value of eq.(\ref{reweight})
one needs extremely high statistics.
Furthermore there is another difficulty: the calculation of the complex phase $\theta$ 
contains a determinant calculation which is computationally costly. 
For small $\mu$ one may use the Taylor expansion of the determinant as\cite{Allton1}
\be
\det M(\mu)= \exp(TrlnM(0)
+\left.Tr\frac1{M}\frac{\partial M}{\partial \mu}\right|_{\mu=0}\times\mu +O(\mu^2)).
\label{Taylor}
\ee 
The first term of eq.(\ref{Taylor}) is real and the second term is pure imaginary.
Thus $\theta$ at small $\mu$ is given by
\be 
\theta  =  \bar{\theta}+O(\mu^3),
\label{theta}
\ee
where
\be
\bar{\theta}=\left.Im Tr\frac1{M}\frac{\partial M}{\partial \mu}\right|_{\mu=0}\times\mu.
\ee
In this expression no determinant calculation is required.
Instead, $\bar{\theta}$ is given by a trace calculation  
and the computational cost is much reduced. 
An empirical study shows that  
the quality of the approximation $\bar{\theta}$ is valid for $\mu \le 0.2$\cite{FKT}.  
Figure~\ref{fig:thetabar} shows $\bar{\theta}$ vs the exact phase $\theta$.
The agreement between $\bar{\theta}$ and  $\theta$ is excellent for $\mu=0.1$ and 0.2. 
However for $\mu=0.3$, $\bar{\theta}$ deviates from the exact value $\theta$.

\begin{figure}
    \vspace{5mm}
       \centerline{\includegraphics[width=6.5 cm]
                                   {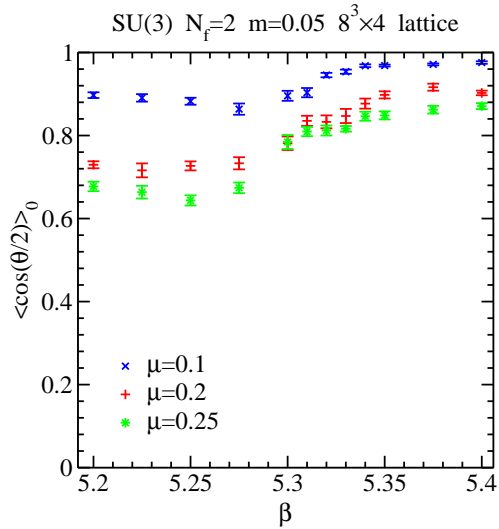}}
   \caption{Phase fluctuation $\la \cos(\theta/2)\ra_{\mu_V}$ for $n_f=2$ 
in the isospin ensemble\cite{SAT}. $\beta_c(\mu=0) \approx 5.32$.}
   \label{fig:COS2}
\end{figure}

For large $\mu$, in order to obtain $\theta$, using the approximation  
is not sufficient and 
one needs the determinant calculation to 
obtain the  value of $\theta$. 
Sasai, Nakamura and Takaishi studied
the phase fluctuations by calculating $\theta$ without any approximation.  
Figure~\ref{fig:COS2} shows the phase fluctuation $\la \cos(\theta/2)\ra_{\mu_V}$ for $n_f=2$ as
a function of $\beta$\cite{SAT}. 
$\la \cos(\theta/2)\ra_{\mu_V}$ is the average value over the phase quenched ensemble.
The phase fluctuation increases, 
i.e., $\la \cos(\theta/2)\ra_{\mu_V}$ decreases as $\mu$ increases.
However in the deconfinement region, the phase fluctuation
is much smaller than  that in the confinement region.

\begin{figure}
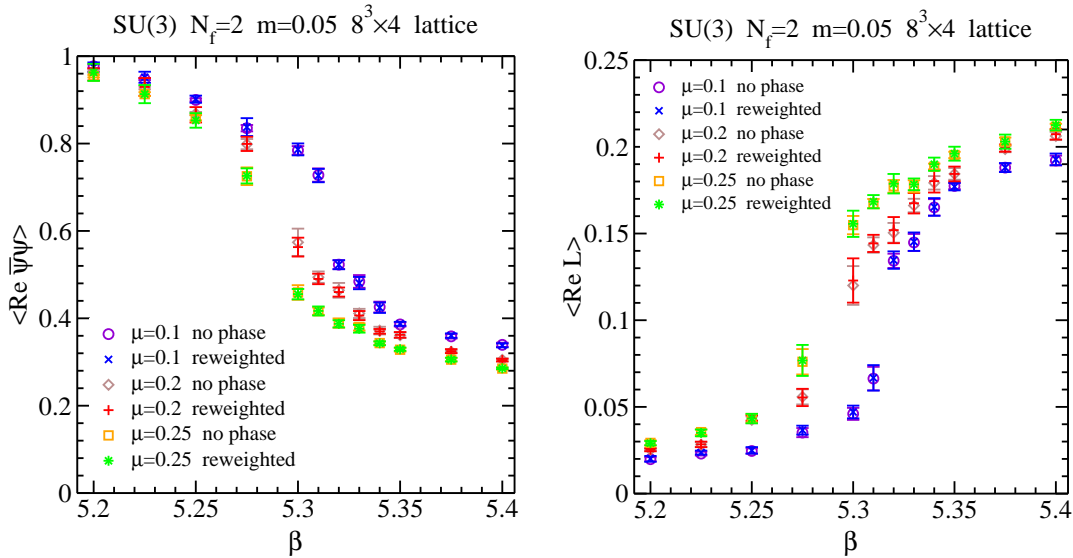

\vspace{3mm}
       \centerline{\includegraphics[width=6.8 cm]
                                   {cpsiR.eps}
\hspace{3mm}
\includegraphics[width=6.8 cm]
                                   {cpolR.eps}}
\caption{$\la\bar{\psi}\psi\ra$ (left) and Polyakov loop (right) 
with and without reweighting\cite{SAT}."no phase" stands for the results without reweighting.}
   \label{fig:REWEIGHTING}
   \end{figure}

With the simulation parameters used here, the phase fluctuation
is not significantly large and 
the expectation values at finite $\mu$ can be obtained by
the reweighting from  the isospin ensemble as eq.(\ref{reweight}).
Figure~\ref{fig:REWEIGHTING} shows $\la\bar{\psi}\psi\ra$ and Polyakov loop with and without reweighting
for  $\mu \leq 0.25$.
We do not see any difference between the results with and without reweighting.
This is consistent with the result of $\la\bar{\psi}\psi\ra$  in \citen{Toussaint}.
This might indicate that  for $\mu \leq 0.25$ the phase effects are small.
This fact is also consistent with that the phase diagram obtained 
at small $\mu_V$ is similar to that at small $\mu_S$\cite{Kogut,NT}.
The phase diagram at small $\mu_V$    
was calculated for $n_f=2$ staggered fermions\cite{Kogut} 
and also for the Wilson fermions\cite{NT} with the plaquette and DBW2\cite{TAKAISHI,TF,TARO2000} gauge actions.

Although the isospin system has a positive measure and can be simulated with the standard Monte Carlo
methods, for large chemical potentials 
we face the computational difficulty: the matrix solver does not converge.
As a result the simulation can not be performed. This feature is also mentioned in \citen{Toussaint}. 
Interestingly for further large chemical potentials $\mu > \mu_c$ 
($\mu_c$ is a certain value which is dependent of the simulation parameters),
the matrix solver converges\cite{SAT2}.

\section{Density of states method at isospin density}
The most lattice simulations are performed by the Monte Carlo methods
based on importance sampling.
Here we report on an alternative approach, the density of states (DOS) method. 
We apply the DOS method for isospin density and
compare the results with those from the standard Monte Carlo method.

The DOS method has been applied for gauge theories\cite{GAUGE}
and QED with dynamical fermions\cite{QED}.
Luo\cite{LUO} applied the DOS method for QCD and argued that
if the eigenvalues of the Dirac operators are determined 
the thermodynamic quantities derived from the partition function can be 
evaluated  at any quark mass and flavor. 

The expectation value of the operator $O$ is given by
\be
\bra O \ket =\frac1{Z}\int [dU] O[U] \det M(m_q,\mu)^{n_f/4} \exp(-\beta S_g[U]),
\label{IMPORTANCE}
\ee
where 
$M(m_q,\mu)$ is assumed to be a staggered fermion matrix at quark mass $m_q$
and at chemical potential $\mu$.
We define  $n(E)$ as
\be
n(E) =\int [dU] \delta(6VE-S_g[U]),
\label{dosn}
\ee
where $V$ is the number of lattice sites and $E$ is the plaquette energy.

Using $n(E)$, $\bra O\ket$ is expressed by
\be
\bra O\ket = \frac1{Z_n}\int  dE n(E) e^{-\beta 6VE}
\bra O\det M(\mu)^{N_f/4} \ket_{E},
\label{dosn2}
\ee
where
\be
Z_n=\int  dE n(E)e^{-\beta 6VE}\bra \det M(\mu)^{N_f/4}
\ket_{E}.
\label{dosn3}
\ee
$\bra \bullet \ket_E$ stands for the microcanonical averages with fixed $E$.
If these microcanonical averages are determined as a function of $E$,
$\la O \ra$ is given at any $\beta$.
Furthermore, Luo argued that if one stores the eigenvalues of $M(\mu)$ for all configurations,
then one can evaluate $\bra \det M(\mu)^{N_f/4} \ket_{E}$
at any quark mass and flavor as
\be
\bra \det M(\mu)^{N_f/4} \ket_{E}
= \bra ( \prod_i^{N_c V}(\lambda_i(\mu)+m_q))^{N_f/4} \ket_E,
\label{det}
\ee
where $\lambda_i(\mu)$ is the i-th eigenvalue of the massless fermion matrix
$M(m_q=0)$ and $N_c=3$ for SU(3).
On the other hand, in general $\bra O\det M(\mu)^{N_f/4} \ket_{E}$ is 
not calculable at any quark mass and flavor.
However for  $O=\bar{\psi}\psi(\mu)$ which is given with 
the trace of $M^{-1}$ one can also evaluate it at any quark mass and flavor as
\be
 \bra (\frac1{V}Tr M(\mu)^{-1}) \det M(\mu)^{N_f/4} \ket_{E}
=\bra \frac1{V}\sum_i \frac1{\lambda_i(\mu)+m_q}
 ( \prod_i^{N_c V}(\lambda_i(\mu)+m_q))^{N_f/4} \ket_{E}.
\label{psidet}
\ee
Using these equations one can obtain $\la\bar{\psi}\psi(\mu)\ra$
at any $\beta$, quark mass and flavor.

\begin{wrapfigure}{r}{7.2cm}
    \vspace{5mm}
        \centerline{\includegraphics[width=7.2 cm]
                                    {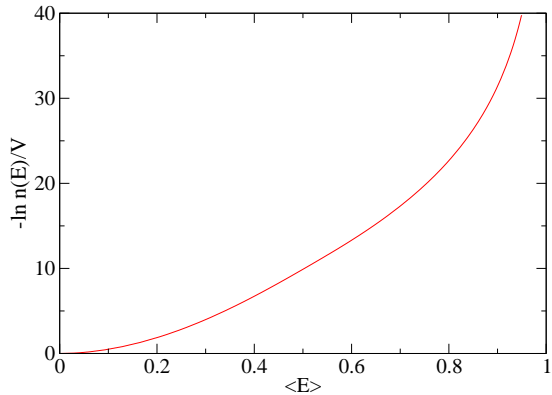}}
   \caption{$-\ln n(E)/V$ as a function of plaquette energy\cite{DOStakaishi}.}
   \label{fig:dos1}
\end{wrapfigure}

Here we comment on the construction of the density of states.
One can define the DOS for any other quantities.
For instance, Gocksch\cite{Gocksch} constructed the DOS
for the complex phase.
Ambjorn {\it et al.}\cite{Ambjorn} constructed 
the DOS for the number density in their factorization method.
In these definitions of the DOS, simulation parameters $\beta$, $n_f$ and 
$m_q$ are absorbed in the DOS and the simulation parameters 
are not variable. 

\begin{figure}
    \vspace{5mm}
        \centerline{\includegraphics[width=7 cm]
                                    {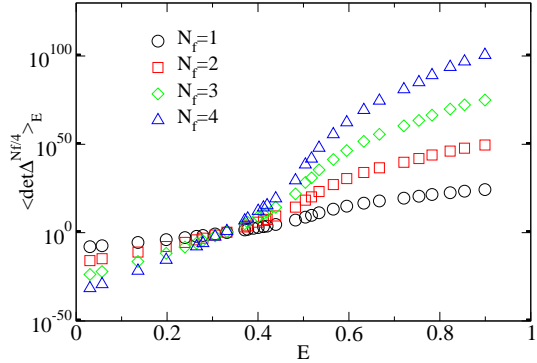}}
   \caption{Microcanonical average of $\bra \det M(\mu_V)^{N_f/4} \ket_{E}$
at $\mu_V=0.2$ and $m_q=0.025$ as a function of $E$\cite{DOStakaishi}
}
   \label{fig:micro}
\end{figure}

In the following we show results on $4^4$ lattices 
at isospin chemical potential $\mu_V$.
The DOS $n(E)$ in eq.(\ref{dosn}) can be obtained using the quenched data as\cite{LUO}
\be
- \frac{\ln n(E)}{V} = 6 \int^E_0 dE^\prime \beta(E^\prime) + const.
\ee
Figure~\ref{fig:dos1} shows $-\ln n(E)/V$ as a function of plaquette energy $E$.
The time consuming part of the method is the calculations of
$\bra \det M(\mu)^{N_f/4} \ket_{E}$ and $\bra O\det M(\mu)^{N_f/4} \ket_{E}$
which contain the eigenvalue calculation.
In order to generate configurations at fixed $E$ we used 
the over-relaxation method.
At each $E$ we generated 100 configurations. Each configuration is separated 
by 100 over-relaxed updates.
For each configuration we calculate eigenvalues and 
those eigenvalues are used to evaluate eqs.(\ref{det}) and (\ref{psidet}).
Figure~\ref{fig:micro} shows $\bra \det M(\mu_V)^{N_f/4} \ket_{E}$
at $\mu_V=0.2$ and $m_q=0.025$ for various flavors as examples. 

\begin{figure}
    \vspace{5mm}
        \centerline{\includegraphics[width=7cm]
                                    {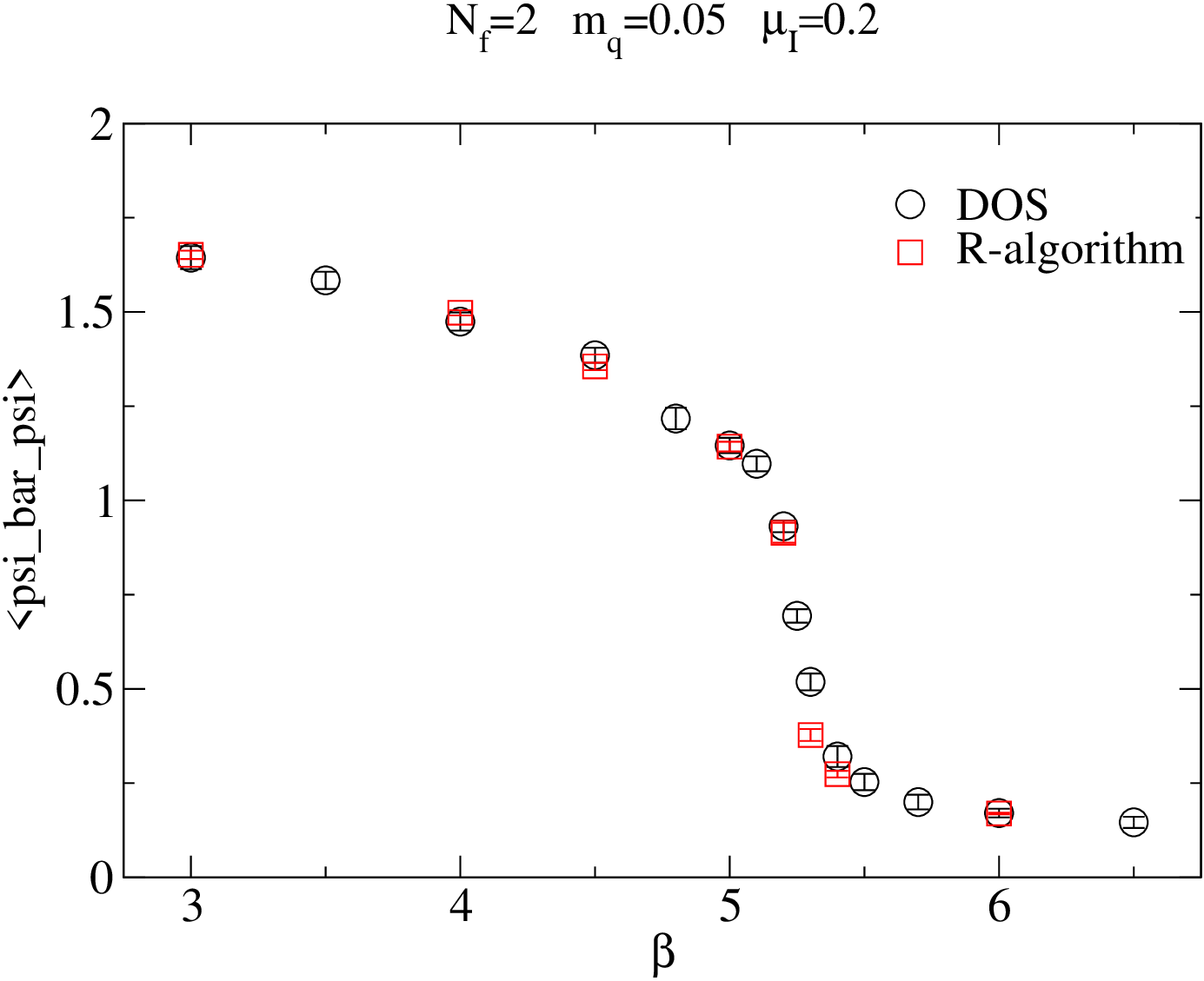}}
   \caption{$\bra \bar{\psi}\psi\ket$ for $n_f=2$ at $\mu_V=0.2$ and $m_q=0.05$\cite{DOStakaishi}.
}  
   \label{fig:DOSvsR}
\end{figure}

The isospin system has a positive measure and can be simulated 
with the standard Monte Carlo algorithm as R-algorithm\cite{R-algo}.
Figure~\ref{fig:DOSvsR} compares the results from the DOS method with 
those from the R-algorithm.
They are in good agreement with each other.
Only a small difference is seen in the phase transition region.   

\begin{figure}
    \vspace{5mm}
        \centerline{\includegraphics[width=7 cm]
                                    {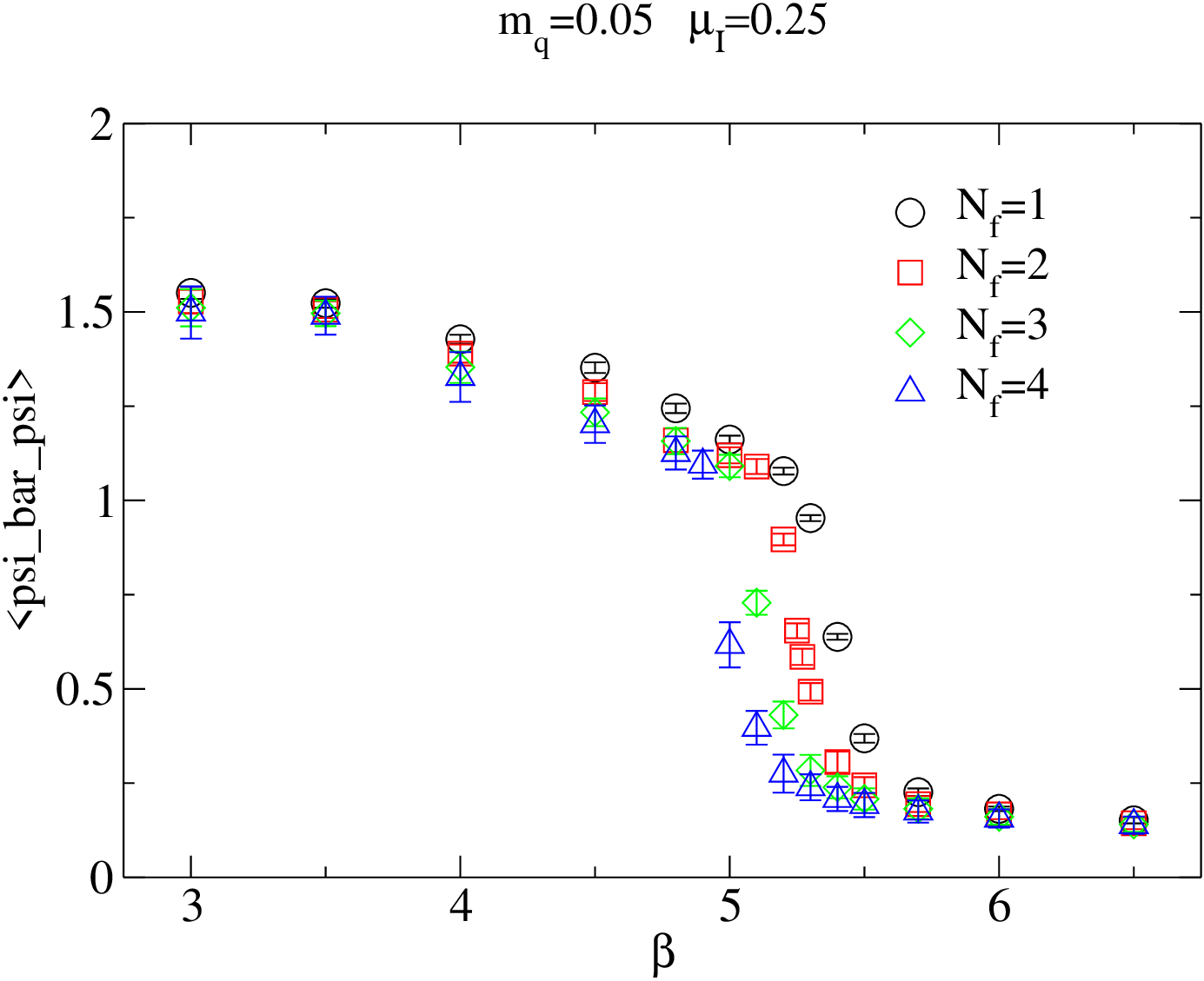}}
   \caption{$\bra \bar{\psi}\psi\ket$ for different $n_f$ at $\mu_V=0.25$ and $m_q=0.05$\cite{DOStakaishi}.
}
   \label{fig:nfall}
\end{figure}

Figure~{\ref{fig:nfall}} shows $\bra \bar{\psi}\psi\ket$ for different $n_f$ at $\mu_V=0.25$ and $m_q=0.05$.
One can see that how $\beta_c$ changes as $n_f$, i.e. $\beta_c$ decreases as $n_f$ increases.

\begin{figure}
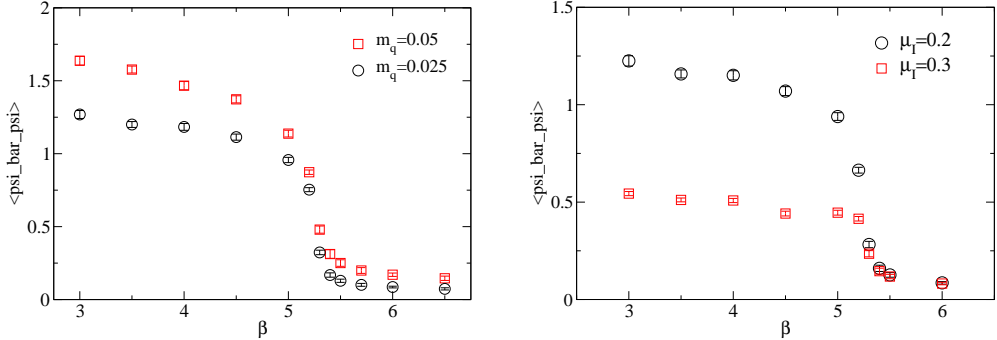

\vspace{5mm}
        \centerline{\includegraphics[width=6.2 cm]
                                    {nf1+nf1mu02mq0025005.eps}
\hspace{5mm}
{\includegraphics[width=6.2 cm]
                                    {mu0.2+mu0.3psiall.eps}}
}
\caption{ 
$\bra \bar{\psi}\psi(m_q,\mu_V)\ket$ for $N_f=1+1$ at $\mu_V=0.2$ with different quark masses,
($m_q=0.05$ and $0.025$)(left), and at $m_q=0.025$ with different chemical potentials ($\mu_V$=0.2 and 0.3)\cite{DOStakaishi}.
}
\label{fig:nondegenerate}
    \end{figure}

In the DOS method one may take various combinations of parameters.
Let us consider the case of $n_f=1+1$ with non-degenerate quark masses $m_1$ and $m_2$.
In this case one must calculate the following microcanonical averages:
\be
\displaystyle \bra |\det \Delta(m_1)|^{N_f/4} |\det \Delta(m_2)|^{N_f/4} \ket_{E},
\ee
\be
\displaystyle \bra  \bar{\psi}\psi(m_{i=1,2}) |\det \Delta(m_1)|^{N_f/4} |\det \Delta(m_2)|^{N_f/4} \ket_{E}.
\ee
Since the eigenvalues are stored, it is easy to calculate these microcanonical averages.
On the other hand, in the conventional algorithm as R-algorithm, it is not easy to simulate the non-degenerate 
system since one needs a different program to perform the simulation.
Figure~\ref{fig:nondegenerate}(left) shows $\bra \bar{\psi}\psi\ket$ for $n_f=1+1$
with different quark masses ( $m_1=0.05$ and $m_2=0.025$ ) at $\mu_V=0.2$.

Similarly one can also consider non-degenerate isospin chemical potentials $\mu_1$ and $\mu_2$.
In this case one calculates
\be
\displaystyle \bra |\det \Delta(\mu_1)|^{N_f/4} |\det \Delta(\mu_2)|^{N_f/4} \ket_{E},
\ee
\be
\displaystyle \bra  \bar{\psi}\psi(\mu_{i=1,2}) |\det \Delta(\mu_1)|^{N_f/4} |\det \Delta(\mu_2)|^{N_f/4} \ket_{E}
.
\ee
Figure~\ref{fig:nondegenerate}(right) shows $\bra \bar{\psi}\psi\ket$ for $n_f=1+1$
with different isospin chemical potentials ( $\mu_1=0.2$ and $\mu_2=0.3$ ) at $m_q=0.025$.

In the DOS method one can easily obtain results for various parameters without making independent simulations.
Thus the DOS method is considered to be useful to explore a wide parameter space. 
On the other hand, the eigenvalue calculations are computationally difficult on large lattices.  
Therefore the application of the DOS method might be limited on small lattices.



\begin{thebibliography}{99}


\bibitem{Review1}
For recent reviews, see e.g.,
S.~Muroya, A.~Nakamura, C.~Nonaka and T.~Takaishi,
Prog.\ Theor.\ Phys.\  {\bf 110} (2003) 615
[arXiv:hep-lat/0306031];
S.~D.~Katz,
arXiv:hep-lat/0310051.


\bibitem{QCD-TARO}
S.~Choe {\it et al.}  [QCD-TARO Collaboration],
Nucl.\ Phys.\ Proc.\ Suppl.\  {\bf 106} (2002) 462
[arXiv:hep-lat/0110223].

\bibitem{QCD-TARO2}
S.~Choe {\it et al.},
Phys.\ Rev.\ D {\bf 65} (2002) 054501; Nucl. Phys. A {\bf 698} (2002) 395.


\bibitem{SUS}
S.Gottlieb {\it et al.}, \PRD{38,1988,2888}

\bibitem{Gavai2003}
R.~V.~Gavai and S.~Gupta,
Phys.\ Rev.\ D {\bf 68} (2003) 034506
[arXiv:hep-lat/0303013].

\bibitem{Allton2003}
C.~R.~Allton, S.~Ejiri, S.~J.~Hands, O.~Kaczmarek, F.~Karsch, E.~Laermann and C.~Schmidt,
Phys.\ Rev.\ D {\bf 68} (2003) 014507
[arXiv:hep-lat/0305007].


\bibitem{PSIBARPSI}
S.Choe, Y.Liu, A.Nakamura and T.Takaishi, in preparation

\bibitem{NJL}
O.~Miyamura, S.~Choe, Y.~Liu, T.~Takaishi and A.~Nakamura,
Phys.\ Rev.\ D {\bf 66} (2002) 077502
[arXiv:hep-lat/0204013].



\bibitem{Allton1}
C.~R.~Allton {\it et al.},
Phys.\ Rev.\ D {\bf 66} (2002) 074507
[arXiv:hep-lat/0204010].

For the full reweighting approach without Taylor expansion, see 
Z.~Fodor and S.~D.~Katz,
Phys.\ Lett.\ B {\bf 534} (2002) 87
[arXiv:hep-lat/0104001];
Z.~Fodor and S.~D.~Katz,
JHEP {\bf 0203} (2002) 014
[arXiv:hep-lat/0106002].


\bibitem{FKT}
P.~de Forcrand, S.~Kim and T.~Takaishi,
Nucl.\ Phys.\ Proc.\ Suppl.\  {\bf 119} (2003) 541
[arXiv:hep-lat/0209126].

\bibitem{SAT}
Y.~Sasai, A.~Nakamura and T.~Takaishi,
arXiv:hep-lat/0310046.


\bibitem{Toussaint}
D.~Toussaint,
Nucl.\ Phys.\ Proc.\ Suppl.\  {\bf 17} (1990) 248.

\bibitem{NT}
A.~Nakamura and T.~Takaishi,
arXiv:hep-lat/0310052.


\bibitem{Kogut}
J.~B.~Kogut and D.~K.~Sinclair,
arXiv:hep-lat/0309042.

\bibitem{TAKAISHI}
T.~Takaishi,
Phys.\ Rev.\ D {\bf 54} (1996) 1050.

\bibitem{TF}
T.~Takaishi and P.~de Forcrand,
Phys.\ Lett.\ B {\bf 428} (1998) 157
[arXiv:hep-lat/9802019].
\bibitem{TARO2000}
P.~de Forcrand {\it et al.}  [QCD-TARO Collaboration],
Nucl.\ Phys.\ B {\bf 577} (2000) 263
[arXiv:hep-lat/9911033].


\bibitem{SAT2}
S.Sasai, A.Nakamura and T.Takaishi, in preparation 

\bibitem{GAUGE}
G.Bhanot, K.Bitar and R.Salvador, Phys. Lett. B187 (1987) 381;
Phys. Lett. B188 (1987) 246;
M.Karliner, S.R.Sharpe and Y.F.Chang, Nucl. Phys. B302 (1988) 204

\bibitem{QED}
V.Azcoiti, G. di Carlo and A.F.Grillo, Phys. Rev. Lett. 65 (1990) 2239
\bibitem{LUO}
X.~Q.~Luo,
Mod.\ Phys.\ Lett.\ A {\bf 16} (2001) 1615
[arXiv:hep-lat/0107013].



\bibitem{Gocksch}
A.~Gocksch,
Phys.\ Rev.\ Lett.\  {\bf 61} (1988) 2054.


\bibitem{Ambjorn}
J.~Ambjorn, K.~N.~Anagnostopoulos, J.~Nishimura and J.~J.~M.~Verbaarschot,
JHEP {\bf 0210} (2002) 062
[arXiv:hep-lat/0208025].


\bibitem{DOStakaishi}
T.~Takaishi,
Mod.\ Phys.\ Lett.\ A {\bf 19} (2004) 909
[arXiv:hep-lat/0312038].


\bibitem{R-algo}
S.Gottlieb, W.Liu, D.Toussaint, R.L.Renken and R.L.Sugar,
Phys. Rev. D {\bf 35} (1987) 2531


\end{thebibliography}
\end{document}